\documentclass{llncs}

\usepackage{url}
\usepackage{multicol}
\usepackage{xspace}
\usepackage{ifpdf}
\ifpdf
    \usepackage[pdftex]{graphicx, color}
\else
    \usepackage[dvips]{graphicx}
\fi

\usepackage{amsmath,amssymb,amsfonts}

\def\p#1{\mathrel{\ooalign{\hfil$\mapstochar\mkern 5mu$\hfil\cr$#1$}}}
\let	\fun		\rightarrow
\def	\pfun		{\p\fun}
\def    \bij            {\rightarrowtail\!\!\!\!\!\rightarrow}

\begin{document}

\title{Refining interfaces: the case of the B method}

\author{David D\'{e}harbe \and Bruno E.G. Gomes \and Anamaria Martins Moreira}

\institute{Universidade Federal do Rio Grande do Norte\\ 
  Departamento de Inform\'{a}tica e Matem\'{a}tica Aplicada \\
  Natal, RN, Brazil
  \email{\{david,bruno,anamaria\}@dimap.ufrn.br}
}

\maketitle

\begin{abstract}
  Model-driven design of software for safety-critical applications
  often relies on mathematically grounded techniques such as the B
  method. Such techniques consist in the successive applications of
  refinements to derive a concrete implementation from an abstract
  specification. Refinement theory defines verification conditions to
  guarantee that such operations preserve the intended behaviour of
  the abstract specifications. One of these conditions requires
  however that concrete operations have exactly the same signatures as
  their abstract counterpart, which is not always a practical
  requirement.  This paper shows how changes of signatures can be
  achieved while still staying within the bounds of refinement
  theory. This makes it possible to take advantage of the mathematical
  guarantees and tool support provided for the current
  refinement-based techniques, such as the B method.
\end{abstract}

\section{Introduction}
\label{sec:introduction}

Java Card~\cite{chen:2000} is a state-of-the-art technology that
provides a programming environment for smart cards that is compatible
with the Java programming language and its underlying platform. Due to
the limited processing power of the chips found on smart cards, Java
Card components are small and require few resources. They thus provide
an interesting testbed for formal approaches to software design such
as the B method~\cite{BBook}. The B method implements a rigorous
model-driven design approach to derive software from a functional
specification through a series of stepwise refinements. It has mature
tool support and has been successfully applied by, e.g. the railway
industry, to develop the software of safety-critical systems.

The goal of the \emph{Bsmart} project~\cite{deharbe06automation} is to
develop a customized version of the B method for the development of
Java Card software components, as well as the corresponding tool
support (as an Eclipse plug-in~\cite{gomes07bsmart}).  Applications
using smart cards have a client-server approach, where the server is a
Java Card component that provides access to the smart card services
and the client is (usually) developed in Java and accesses such
services through a mechanism such as a remote method
invocation. Although based on the same programming paradigms, the type
system of Java Card is much simpler and restricted than that of Java.
Java client software often requires services in a richer type system
than that provided by the Java Card services, and the APIs need to be
adapted. So, in order to be able to include a richer type system in
Bsmart, it appears necessary to include a refinement step
corresponding to such interface adaptation.

Unfortunately, the concept of refinement used in the B method does not
allow for modification of the signature of the operations that compose
such interfaces~\cite{BBook}. Retrenchment~\cite{Banach1ret} is a much
more flexible concept of model transformation, that includes changes
in signature operations and, consequently, in component interfaces.
The scope of retrenchment is however much larger than simple interface
changes, and also includes handling much deeper model transformations,
such as, e.g.  strengthening pre-conditions of operations. This extra
flexibility allows implementations that exhibit behaviours that are
not in the original functional specification, which may not be
desirable in a rigorous model-based development. Also, although the
proponents of retrenchment have developed syntactic extensions to the
B method to include such transformation, these extensions do not yet
benefit from the same level of tool support as refinement.

The goal of this paper is to show a solution to interface changes that
fits within the classical theory of refinement. Thus, it does not
require employing retrenchment and introducing model transformations
that result in executions that are not modeled in the initial
functional specification. In addition, the solution proposed in this
paper consists in model transformations that are fully compatible with
existing tool support for the B method. Indeed, we have defined the
generic refinement pattern, as well as an instance thereof, in B
itself and have used existing tools to prove their
correctness. 

Several authors have related interface changes with
refinement~\cite{hayes95specification,mikhajlova97class,stepney98more,boiten98iorefinement},
however none of thes works is related to the B method; also they change
the verification conditions associated to
refinement. In~\cite{colin09trustworthy}, an approach similar to ours
is presented in the context of component-based development; however
they do not go so far as to present a refinement pattern as detailed
as the one presented in this paper.

\paragraph{Plan of the paper.} Section~\ref{sec:b-method} briefly
introduces the B method and introduces an example that will be used
throughout the paper to illustrate the different model
transformations. Also, the main concepts of retrenchments are exposed
and discussed in
Section~\ref{sec:retrenchment}. Section~\ref{sec:interface} then
presents the refinement pattern to introduce interface changes and a
model transformation instantiating this pattern is presented in
Section~\ref{sec:case-study}. Finally, conclusions and future work are
presented in Section~\ref{sec:conclusions}.
 
\section{Model-driven development with B}
\label{sec:b-method}

The B method for software development \cite{BBook,bref:2007} is a
model-driven development method based on formal models and formally
verified derivations or refinements. It provides the B {\em Abstract
  Machine Notation} (AMN) to represent models at different levels of
abstraction, based on first order logic, integer arithmetic and set
theory. These different levels of abstraction of a model must be
related by formally proved refinements.

Industrial tools for the development of B based projects have been
available for a while now \cite{atelierb,BToolkit}, with specification
and verification support as well as some project management tasks and
support for team work. More recently, various academic and/or open
source tools have spread, and Atelier B \cite{atelierb} has become
free of charge, increasing the popularity of the method and the
variety of its uses.

\subsection{The B development process}

A B specification is structured in components. The initial model from
which the software development process initiates may be modularly
composed of one or more {\em MACHINE}s. Such models must be proved
satisfiable (i.e. that they have an implementation) and consistent
with respect to some specified properties (namely, the INVARIANT of
each MACHINE).

Once an abstract model is proved consistent, it may be used as input
for a series of (optional) refinements. The result of each refinement
step for each MACHINE is a new (usually less abstract) module
classified as a {\em REFINEMENT}.  The obtained refined model is
then proved correct with respect to the abstract model. This is done
modularly, by proving the correctness of each REFINEMENT component
with respect to its corresponding machine and to all intermediate
REFINEMENT components in between the abstract MACHINE and the
REFINEMENT being verified.

Eventually, a final refinement takes place, which gives origin to a B
\emph{IMPLEMENTATION}, a special kind of refinement from which code in
a programming language can be generated. The verification of the model
at the IMPLEMENTATION level is carried out similarly as for
refinements, with the addition of the so-called {\em B0 check}, which
is responsible for verifying that the constructs in each
IMPLEMENTATION module are compatible with the used code generator.

Finally, B IMPLEMENTATIONs are used as input for code generation
in some programming language (e.g., C, Ada or Java).  If all
verifications were discharged, and assuming the correctness of the
code generator, this generated code satisfies the stated properties of
the abstract model.

\subsection{Components of a model in the B notation}

A B component contains two main parts: a state space definition and a
set of transitions. The state space is specified as a logic formula
called the invariant. Transitions are specified by means of
\emph{operations}; generally, each operation may take arguments and
return results corresponding to a desired functionality of the
system. The set of initial states is specified as a special operation
(without parameters nor results) called the \emph{initialisation}. A B
component may additionally contain clauses in many forms (parameters,
constants, assertions). Such clauses are not essential in the B
language, but are useful to make specifications and proofs shorter or
more readable.

The specification of the state components appears in the VARIABLES and
INVARIANT clauses.  The former enumerates the state components, and
the latter defines restrictions on the possible values they can take.

For the specification of a module's operations, B offers a language of
so-called {\em generalized substitutions}, ``imperative-like''
constructions with translation rules that define their semantics as
the effect they have on the values of any expression on the (global or
local) variables to which they are applied. The semantics of the
substitutions is defined by the \emph{substitution calculus}, a set of
rules stating how the application of the different forms substitution
rewrite to formulas in first-order logic. Let $S$ denote a
substitution, $E$ an expression, then $\lbrack S \rbrack E$ denotes
the result of applying $S$ to $E$.

Operations are composed of a pre-condition $P$ and a substitution $S$.
Syntactically, this is expressed as $\mathbf{PRE}\, P\,
\mathbf{THEN}\, S\, \mathbf{END}$. In this construct, $P$ specifies
the bounds of application of the operation, and $S$ specifies what
transformations will be applied to the state, as well as how the
operation results (if any) are computed. Operations also have optional
parameters and results. The pre-condition $P$ must establish at least
typing constraints on the parameters and the substitution $S$ define
the value of the results. To establish that an operation does not
drive the component from a valid state to an invalid state, one must
show that the operation, whenever applied in a state that satisfies
the pre-condition, maintains the invariant, i.e.
$$I \land P \Rightarrow \lbrack S \rbrack I.$$

The simplest substitution in the B language is $v := E$ where $v$ is a
variable and $E$ denotes some expression. The semantics is defined as:

\begin{eqnarray*}
\lbrack v := E \rbrack P & \Leftrightarrow & P \langle v \leftarrow E
\rangle,
\end{eqnarray*}

i.e. all free occurrences of $v$ in $P$ are replaced by $E$.

Another substitution that is used in the rest of the paper is a form of
non-deterministic assignment $v :\in V$, where $v$ is allowed to take any
value in the set $V$. The semantics is:
\begin{eqnarray*}
\lbrack v :\in V \rbrack P & \Leftrightarrow & 
\forall x \bullet (x \in V \Rightarrow P \langle v \leftarrow x\rangle),
\end{eqnarray*}

where $x$ is a fresh variable. Note that, for such substitution to be
well-defined, one must show that $V$ is not an empty set.

\subsection{Example of a B Machine}
\label{sec:counter}

In this section, we present a simple example of a B model that will be
used throughout the paper.  Our example is that of a simple counter
(Figure \ref{fig:counter}). In the next sections, this abstract
specification, which intends to specify the Application Programming
Interface (API) of a counter service, will be refined with the
intention of having this service offered by a Smart Card running Java
Card. Some difficulties in this process motivate our proposal.

\begin{figure}[ht!]
\begin{small}
\bf MACHINE\hspace*{0.20in}\it JCounter

\bf SEES \hspace*{0.20in}\it JInt

\bf VARIABLES \hspace*{0.20in}\it value
\hspace*{0.20in}

\bf INVARIANT \hspace*{0.20in}\it value  $\in$  \it JINT
\hspace*{0.20in}

\bf INITIALISATION \hspace*{0.20in}\it value \rm := \it jint\_of \rm(0)
\hspace*{0.20in}

\bf OPERATIONS

\hspace*{0.20in}\bf increment \rm (\it vv\rm )\rm = 

\hspace*{0.20in}\bf PRE \hspace*{0.20in}\it vv  $\in$  \it JINT  $\land$ 
\it sum\_jint\rm (\it value\rm , \it vv\rm ) $\in$  \it JINT 

\hspace*{0.20in}\bf THEN \hspace*{0.20in}\it value \rm :=  \it sum\_jint\rm (\it value\rm , \it vv\rm )

\hspace*{0.20in}\bf END\rm ;

\hspace*{0.20in}\bf decrement \rm (\it vv\rm ) \rm = 

\hspace*{0.20in}\bf PRE \hspace*{0.20in}\it vv  $\in$  \it JINT  $\land$ 
\it subt\_jint\rm (\it value\rm , \it vv\rm ) $\in$  \it JINT  $\land$ 
\rm (\it value \rm - \it vv\rm )  $\geq$  \rm 0

\hspace*{0.20in}\bf THEN \hspace*{0.20in}\it value  \rm :=  \it subt\_jint\rm (\it value\rm , \it vv\rm )

\hspace*{0.20in}\bf END\rm ;

\hspace*{0.20in}\it cc  $\leftarrow$  \bf getCounterValue \rm = \it cc \rm := \it value

\bf END
\end{small}
\caption{The counter machine}
\label{fig:counter}
\end{figure}

In the \emph{JCounter} machine, the variable \emph{value} is the state
component that stores the actual value of the counter.  This variable
is typed as \emph{JINT}, an integer set defined in the \emph{JInt}
machine that corresponds to, e.g., the \emph{int} type of Java
language (this machine belongs to a library of B models, under
development by our group, to support the formal development of Java
and Java Card software).  The machine \emph{JInt}, not shown in the
paper, also defines arithmetic operations on this set so that they
operate within the range for a Java value of type \emph{int}. The
included \emph{JInt} machine uses these properties in the definition
of functions that can be used in substitution of the B operators in
the body of an operation. The specification \emph{JCounter} comprises
three operations to increment, decrement and query the counter.  In
the last operation, the pre-condition is omitted, which is interpreted
as a trivial pre-condition (i.e. $\mathit{TRUE}$). Note that, in the
body of the operations, we use the arithmetic functions defined in
\emph{JInt} machine instead of the B operators for integers. This
means that we could be talking of any other type of data not directly
available in B.

\subsection{Refinements in B}

Refinements, which are central to the proposal of this paper, play a
very important role on the B method. They are responsible for the
creation of a hierarchy of models where each model is proved to be
compliant, according to the B refinement rules \cite{BBook}, to the
previous (more abstract) one in the chain. We briefly present these
rules in the following.

\begin{enumerate}

\item exactly the same number of operations
\item exactly the same operation interfaces (names, parameters and results)
\item each concrete operation must satisfy the classical rules stating that:

\begin{enumerate}
\item it must be applicable whenever its abstract counterpart is (the
  satisfaction of the precondition of the abstract operation must lead
  to the satisfaction of the precondition of the concrete operation).
 
\item when the abstract operation is applicable and the concrete
  operation is applied instead, the observed behavior must be
  compatible with one of the behaviors specified in the abstract
  operation.

\end{enumerate}

\end{enumerate}

\section{Retrenchment}
\label{sec:retrenchment}

The refinement rules presented in the previous section aim to
guarantee that any implementation of the concrete model can be
transparently used as an implementation of the abstract model, but
they are sometimes considered an unnecessary burden to refinement
based development~\cite{Banach1ret}. Refinement rules can hinder its
adoption on the design of a wide range of real world applications, as
the differences between an elegant abstract model and a concrete
model, where implementation needs begin to show up, may not fit the
refinement framework. If we consider, for instance, our counter
example of section \ref{sec:counter} and the need to implement it in a
platform where only short integers are available (this may happen in
some smart cards), there will be a problem with the operations'
interfaces, which are supposed to communicate regular length integers.
Changing the abstract specification to make the development fit in the
refinement framework is not a good approach, as it degrades
reusability and requires verifying the abstract level
again. Retrenchment and the approach presented in this paper do not
need any changes in the abstract specification.

With the main motivation of extending the applicability of formal
development techniques to a wider range of applications, Banach and
Poppleton have proposed a technique called
\emph{retrenchment}~\cite{Banach1ret}, a formal approach to
model-driven design that imposes less constraining rules than
refinement.

Indeed, with retrenchment it is possible to have stronger
preconditions and/or weaker post-conditions in an operation, to change
an operation interface and to transfer behavior from state components
to I/O or vice-versa.  In \cite{Banach1ret} the authors present the
theory and its applicability and demonstrate how to incorporate it as
an extension of the B method. In the following we present a brief
explanation of this extension, concentrating on the possibility to
change an operation's interface during the development process, the
feature which is the focus of this paper.

\subsection{Retrenchment in B}

In Banach and Poppleton proposal, a retrenchment is a B machine with
the addition of: (1) a RETRIEVES clause, to specify the retrieval
relation, relating abstract and concrete variables\footnote{Unlike B
  refinements, where the local invariant and the relation between the
  abstract and concrete states (retrieve relation) are both specified
  in the INVARIANT predicate, in a retrenchment module, the INVARIANT
  only specifies the more concrete state variables. The relation
  between the retrenching and the retrenched states is placed in the
  RETRIEVES clause.}, and (2) {\em ramified generalized
  substitutions}, constructed with the clauses LVAR, WITHIN and
CONCEDES, which extend each operation's generalized substitution,
specifying the situations where the concrete operation fails to refine
the abstract one (Figure~\ref{fig:retrenchment-mch}).

All the elements to describe a B refinement are available to define a
retrenchment: for instance, set definitions clause (SETS) and all the
clauses for machine composition (SEES, INCLUDES, USES, PROMOTES and
EXTENDS) can be used as in a traditional B module.

\begin{figure}[ht!]
$$
\begin{array}[t]{ll}
\begin{array}[t]{l}
\mathbf{MACHINE} \quad \mathit{M}_A (pm_M) \\
\mathbf{CONSTRAINTS} \quad P_A (pm_A)\\
...\\
\mathbf{VARIABLES} \quad v_A  \\
\mathbf{INVARIANT} \\
\quad Inv (v_A) \\
\mathbf{INITIALISATION} \\
\quad Init (v_A) \\
\mathbf{OPERATIONS} \\
\quad r_A \longleftarrow \mathit{OP}_A (p_A) = \\
\quad \quad S_A (v_A, p_A, r_A) \\
\mathbf{END}
\end{array}
&
\begin{array}[t]{l}
\mathbf{MACHINE} \quad \mathit{M}_R (pm_R) \\
\mathbf{CONSTRAINTS} \quad P_R (pm_R) \\
...\\
\mathbf{VARIABLES} \quad v_R  \\
\mathbf{INVARIANT} \\
\quad Inv (v_R) \\
\mathbf{RETRIEVES}\\
\quad Ret (v_A, v_R)\\
\mathbf{INITIALISATION} \\
\quad Init (v_R) \\
\mathbf{OPERATIONS} \\
\quad r_R \longleftarrow \mathit{OP}_R (p_R) = \\
\quad \mathbf{BEGIN}\\
\quad \quad S_R (v_R, p_R, r_R) \\
\quad \mathbf{LVAR}\\
\quad \quad R\\
\quad \mathbf{WITHIN}\\
\quad \quad W (p_A, p_R, v_A, v_R, R)\\
\quad \mathbf{CONCEDES}\\
\quad \quad C (v_A, v_R, r_A, r_R, R)\\
\mathbf{END}
\end{array}
\end{array}
$$
\caption{classical B machine (left) and retrenchment machine (right)}
\label{fig:retrenchment-mch}
\end{figure}


The LVAR clause is optional, and may be used to declare variables
whose scope is the WITHIN and CONCEDES clauses. When present, these
variables must be typed and restricted in the WITHIN clause, which may
also strengthen the operation's precondition. The CONCEDES clause in
turn possibly weakens the post-condition of the operation.

The role of these additional clauses can be more precisely described
through the definition of retrenchment proof obligations, presented in
the following section. Then, in Section~\ref{sec:retsample}, we use a
small example to illustrate how retrenchment works in practice.

\subsubsection{Retrenchment proof obligations}

Retrenchment proof obligations can be classified as local proof obligations,
when only dealing with local data, and joint proof obligations, when 
addressing both the retrenched and retrenching components.

The local proof obligations of a retrenchment module are the same as
those for a regular B machine: establishment of the invariant by the
initialisation; preservation of the invariant by the operations, when
applied to states where their preconditions are satisfied. By
discharging these obligations, one guarantees the internal consistency
of the module.

The joint proof obligations concern initialization and operations. The
initialisation joint proof obligation is similar to that of a
refinement, except for the fact that it is the satisfaction of the
RETRIEVES predicate, instead of the INVARIANT, that is checked:

$$P_A(pm_A) \land P_R(pm_R) \Rightarrow [Init(v_R)] \lnot[Init(v_A)] \lnot Ret(v_A, v_R)$$

The proof obligations for the operations are the most relevant to the
retrenchment framework.  For each operation, correctness verification
is conditioned to situations where: regular B constraints ($P_A(pm_A)$
and $P_R(pm_R)$), abstract and concrete invariants ($Inv(v_A)$ and
$Inv(v_R)$) and the retrieve relation ($Ret(v_A, v_R)$) are satisfied;
the concrete operation terminates ($trm(S_R(v_R, p_R, r_R))$) and the
conditions stated on the WITHIN clause $W(p_A, p_R, v_A, v_R, A)$ are
also satisfied.  The first conditions are similar to those in a
refinement proof obligation. It is important to notice that,
differently from refinement, which requires correctness in each
situation where the abstract operation terminates, it is the
termination of the concrete operation that conditions the
verification.

On the other hand, on the right hand side, we have the option of not
satisfying the retrieve relation (i.e., having a concrete behaviour
which does not correspond to a specified abstract behaviour) as long
as the predicate in the CONCEDES clause is satisfied.

\begin{multline*}
P_A(pm_A) \land P_R(pm_R) \land (Inv(v_A) \land Ret(v_A, v_R) \land Inv(v_R)) \land \\
trm(S_R(v_R, p_R, r_R)) \land W(p_A, p_R, v_A, v_R, R)) \Rightarrow trm(S_A(v_A, p_A, r_A)) \land \\ 
[S_R(v_R, p_R, r_R)] \lnot [S_A(v_A, p_A, r_A)] \lnot (Ret(v_A, v_R) \lor C
(v_A, v_R, r_A, r_R, R))
\end{multline*}

\subsubsection{Retrenching JCounter}\label{sec:retsample}

In this section, we apply retrenchment in the formal development of
the counter service of section \ref{sec:counter} for a version of the
Java Card platform without support for the Java type \emph{int}
(32-bit integers).
 
The architecture of a Java Card application is composed by host-side
software and server-side software. The host side is developed in
standard Java and requests the services supplied by the server
application, called \emph{applet}. The latter resides inside the smart
card chip, which provides a computer with limited memory resources and
processing power. Moreover, the Java Card language is much more
limited than Java (for instance, it has a smaller set of basic types).

We assume that the smart card will provide the token counter service,
that will be used by host-side applications written in Java. The
development starts with the specification of the Java API that will be
available to host-side clients (Figure~\ref{fig:counter}). The
obtained retrenchment, with different operation signatures than those of the
abstract machine, is shown in Figure \ref{fig:jcounterRet}.

\begin{figure}[ht!]
\bf MACHINE \it JCounter\_ret

\bf RETRENCHES \it JCounter

\bf SEES \it JInt,  JCInt, InterfaceContext

\bf VARIABLES \it cvalue

\bf INVARIANT \it cvalue  $\in$  \it JCInt

\bf RETRIEVES \it value \rm = \it jint\_of\_jcint \rm (\it value\rm)

\bf INITIALISATION \it cvalue \rm := \rm \it jcint\_of \rm(0)

\bf OPERATIONS

\hspace*{0.20in}\bf increment \rm ( \it cvv \rm ) \rm =

\hspace*{0.20in}\bf BEGIN

\hspace*{0.40in}\bf PRE \it cvv  $\in$  \it JCINT  $\land$ 
\it sum\_jcint\rm (\it cvalue\rm , \it cvv\rm ) $\in$  \it JCINT 

\hspace*{0.40in}\bf THEN \it cvalue \rm :=  \it sum\_jcint\rm (\it cvalue\rm , \it cvv\rm )

\hspace*{0.40in}\bf END\rm
 
\hspace*{0.20in}\bf WITHIN \it vv \rm = \it jint\_of\_jcint \rm(\it cvv\rm)\rm 

\hspace*{0.20in}\bf END;

\hspace*{0.20in}\bf decrement \rm ( \it cvv \rm ) \rm =

\hspace*{0.20in}\bf BEGIN

\hspace*{0.40in}\bf PRE \it cvv  $\in$  \it JCINT  $\land$ 
\it subt\_jcint\rm (\it cvalue\rm , \it cvv\rm ) $\in$  \it JCINT  $\land$ 

\hspace*{0.60in}\bf \it subt\_jcint\rm (\it cvalue\rm , \it cvv\rm ) $\geq$ 0

\hspace*{0.40in}\bf THEN \it cvalue \rm :=  \it subt\_jcint\rm (\it cvalue\rm , \it cvv\rm )

\hspace*{0.40in}\bf END\rm
 
\hspace*{0.20in}\bf WITHIN\rm 

\hspace*{0.40in}\bf \it vv \rm = \it jint\_of\_jcint \rm(\it cvv\rm)\rm 

\hspace*{0.20in}\bf END;  

\hspace*{0.20in}\it cc  $\leftarrow$  \bf getCounterValue \rm = 

\hspace*{0.20in}\bf BEGIN

\hspace*{0.40in} \it ccc \rm = \it cvalue \rm
 
\hspace*{0.20in}\bf CONCEDES\rm 

\hspace*{0.40in}\bf \it cc \rm = \it jint\_of\_jcint \rm(\it ccc\rm)\rm 

\hspace*{0.20in}\bf END

\bf END
\caption{A retrenchment of JCounter}
\label{fig:jcounterRet}
\end{figure}

As in our example, the basic data type \emph{int} is not available,
one possible solution is to represent it as a combination of available
types, such as \emph{short}. This representation is defined in the
\emph{JCInt} component (not shown) which defines the \emph{JCINT}
type, operators such as addition ({\it sum\_jcint}) and subtraction
({\it subt\_jcint}), and a type cast operation ({\it jcint\_of}) to
generate \emph{JCINT} values from regular B integer values. The
operations in \emph{JCounter\_ret} machine of Figure
\ref{fig:jcounterRet} run completely on this domain. This can be seen
when observing the substitutions that specify the behaviour of each
operation. 

\emph{JCounter\_ret} also imports, through the {\bf SEES} construct,
\emph{JINT} and \emph{InterfaceContext} (described in
Section~\ref{sec:case-study}), which, as one can see, only appear in
the clauses related to retrenchment where they are used to specify the
relation between specifications (\emph{JCounter} and
\emph{JCounter\_ret}). {\it jint\_of\_jcint} is a bijection, defined
in \emph{InterfaceContext} associating each Java integer to its Java
Card representation. It is used in four different places: to specify
the retrieve relation as it would regularly be done in a refinement;
and in each operation, to associate each concrete parameter or result
to its abstract counterpart. In a refinement, because there can be no
changes in interfaces, this association is done automatically and does
not need to be stated.

\subsection{Some Notes on Retrenchment}


Although retrenchment could be an attractive alternative to strict
refinement for some developments, its adoption is currently not
expressive and there is not yet a mature tool support for it. An
academic initiative in this direction is the Frog
tool~\cite{frogToolkit}, developed as part of the PhD thesis of
Frasier~\cite{FraserRet08}. The tool proposes a framework to mechanize
the support for retrenchment. Initially the Z~\cite{Z} notation was used as
mathematical notation and the proof obligations were generated to the
Isabelle theorem prover. But as the proposal of the framework is to be
extensive, one can use it to configure its own formal model based
development.

In the next section of the paper, we describe our solution to the
problem of interface adaptation and type changing between models
without going out the refinement theory using the B method.

\section{Interface adaptation as refinement}
\label{sec:interface}

This section describes a way how model transformations consisting of a
modification in the signature of operations, can be performed by means
of refinement. This transformation is presented as a refinement
pattern~\cite{Lecomte07patrons} written and developed with the B
method itself. Such pattern will then be instantiated in
Section~\ref{sec:case-study} for a simple software development for the
Java Card platform.

\subsection{A schematic specification in B}

We first present the schema of a specification model in B. This schema
is described in the B language itself as a component named
$\mathit{API}_A$, that is presented in
Figure~\ref{fig:component-api}. The types, sets and relations employed
in the machine $\mathit{API}_A$ are defined in the component
$\mathit{Context}_A$, presented in
Figure~\ref{fig:component-context}. Note that, for the sake of
conciseness, the $\mathit{API}_A$ machine only includes the clauses
that provide the essence of what is a B model, namely a set of states,
constrained by an invariant predicate, a set of transitions and
initial states, both specified by means of substitutions. So, while
there is no parameters, constants and sets in this pattern machine,
the generality of the approach is thus not compromised.

A B component modelling a system has a state, and it is represented
here as a single variable $v_A$, of type $\mathit{type}_A$ (defined in
$\mathit{Context}_A$). The valid states are identified by the set
$\mathit{inv}_A$ and the initial states by the set $\mathit{init}_A$.

The transitions of the system are modelled by a single operation,
named $\mathit{operation}_A$.  The parameters of the operations are
represented by $\mathit{p}_A$ and its results by $\mathit{r}_A$. In
the general case, an operation may have a precondition that depends on
the state and parameters.  It is here specified by means of the set
$\mathit{pre}_A$. The possible next states and output values are
chosen non-deterministically amongst the sets of values denoted
$\mathit{stf}_A$ and $\mathit{ouf}_A$ respectively; both depend on the
state variable and the operation parameter.

\begin{figure}
$$
\begin{array}{l}
\mathbf{MACHINE} \quad \mathit{API}_A \\
\mathbf{SEES} \quad \mathit{Context}_A \\
\mathbf{VARIABLES} \quad v_A  \\
\mathbf{INVARIANT} \\
\quad v_A \in \mathit{type}_A \land v_A \in \mathit{inv}_A \\
\mathbf{INITIALISATION} \\
\quad v_A :\in \mathit{init}_A \\
\mathbf{OPERATIONS} \\
\quad r_A \longleftarrow \mathit{operation}_A (p_A) = \\
\quad \mathbf{PRE} \quad p_A \in \mathit{type}_A \land (v_A, p_A) \in \mathit{pre}_A \quad \mathbf{THEN} \\
\quad \quad v_A :\in \mathit{stf}(v_A, p_A) \parallel r_A :\in \mathit{ouf}_A(v_A, p_A) \\
\quad \mathbf{END} \\
\mathbf{END}
\end{array}
$$
\caption{A pattern for an abstract specification in B}
\label{fig:component-api}
\end{figure}

In order to be able to prove the validity of the verification
conditions of the component $\mathit{API}_A$, the objects defined in
$\mathit{Context}_A$ need to satisfy a number of constraints, that are
stated in its $\mathbf{PROPERTIES}$ clause.

The first five constraints are typing conditions, the next two
constraints state that the domain of the state transition and output
relations must contain the valid states and operation parameters. The
last two constraints must be also satisfied to guarantee that all the
reachable states of the component are also valid states (i.e. in the 
set representing the invariant).

\begin{figure}
$$
\begin{array}{l}
\mathbf{MACHINE} \quad \mathit{Context}_A \\
\mathbf{SETS} \quad \mathit{type}_A \\
\mathbf{CONSTANTS} \\
\quad \mathit{stf}_A, \quad \quad \quad \mbox{(* state transition function *)} \\
\quad \mathit{ouf}_A, \quad \quad \quad \mbox{(* output function *)} \\
\quad \mathit{inv}_A, \quad \quad \quad \mbox{(* state invariant *)} \\
\quad \mathit{init}_A, \quad \quad \quad \mbox{(* initial states *)} \\
\quad \mathit{pre}_A \quad \quad \quad \mbox{(* operation precondition: depends on state and parameter *)} \\
\mathbf{PROPERTIES} \\
\quad \mathit{inv}_A \subseteq \mathit{type}_A \quad \land \\
\quad \mathit{init}_A \subseteq \mathit{type}_A \quad \land \\
\quad \mathit{pre}_A \subseteq \mathit{type}_A \times \mathit{type}_A \quad \land \\
\quad \mathit{stf}_A \subseteq (\mathit{type}_A \times \mathit{type}_A) \leftrightarrow\mathit{type}_A \quad \land \\
\quad \mathit{ouf}_A \subseteq (\mathit{type}_A \times \mathit{type}_A) \leftrightarrow \mathit{type}_A \quad \land \\
\quad \mathit{inv}_A \lhd \mathit{pre}_A \subseteq \mathbf{dom}(\mathit{stf}_A)  \quad \land \\
\quad \mathit{inv}_A \lhd \mathit{pre}_A \subseteq \mathbf{dom}(\mathit{ouf}_A)  \quad \land \\
\quad \mathit{init}_A \subseteq \mathit{inv}_A \quad \land \\
\quad \mathit{stf}_A \lbrack \mathit{inv}_A \lhd \mathit{pre}_A \rbrack \subseteq \mathit{inv}_A \\
\mathbf{END}
\end{array}
$$
\caption{Component defining the objects used in component $\mathit{API}_A$}
\label{fig:component-context}
\end{figure}

Assume now that the component $\mathit{API}_{A}$ is to be refined by a
component $\mathit{API}_C$ such that the data carried by state
variables, operations parameters and results may be different. In the
following, the objects in the component $\mathit{API}_C$ will be here
designated as the objects in the component $\mathit{API}_A$, with the
$A$ subscript substituted by the $C$ subscript. For instance, the
signature of the operation in the machine $\mathit{API}_C$ is:
$$
    r_C \longleftarrow \mathit{operation}_C (p_C) =
$$ where $p_C$ satisfies $p_C \in \mathit{type_C} \land p_C \in
\mathit{pre}_C$.  Since the refinement of operations must preserve
their signature, it is necessary to propose a workaround, such as
retrenchment does. In the next section we show a refinement pattern
that makes it possible to use operations with a different signature in
a refinement.

\subsection{A refinement pattern for signature changes}

The main idea that underlies the pattern is to use an interface
adapter (see Figure~\ref{fig:refinement-pattern}).  Note that this
refinement is solely responsible for interfacing the two components
and is not meant to introduce other design decisions such as reducing
non-determinism, or precondition weakening. 

The $\mathit{API}_A$ component is refined by a component
$\mathit{API}_r$.  This refinement includes an instance of the
component $\mathit{API}_C$, and the \emph{gluing} invariant
establishes the relationship between the state of $\mathit{API}_A$ and
the state of $\mathit{API}_C$.  In $\mathit{API}_r$, the operation has
the same signature as in $\mathit{API}_A$. It consists of a three-step
sequence. First the value of the parameter $p_A$ is translated to
corresponding value of type $\mathit{type}_C$ and the result is stored
in variable $\mathit{to}$. Second, $\mathit{operation}_C$ is applied
to $\mathit{to}$ and the result is stored in a variable
$\mathit{from}$. The value of $\mathit{from}$ is then converted back
to $\mathit{type}_A$ and returned.

\begin{figure}
$$
\begin{array}{l}
\mathbf{REFINEMENT} \quad \mathit{API}_r \\
\mathbf{REFINES} \quad \mathit{API}_A \\
\mathbf{SEES} \quad \mathit{Context}_A, \mathit{Context}_C, \mathit{Context}_I \\
\mathbf{INCLUDES} \quad \mathit{API}_C \\
\mathbf{INVARIANT} \quad v_A = \mathit{AofC} (v_C) \\
\mathbf{OPERATIONS} \\
\quad r_A \longleftarrow \mathit{operation}_A (p_A) = \\
\quad \mathbf{VAR} \quad \mathit{to}, \mathit{from} \quad \mathbf{IN} \\
\quad \quad \mathit{to} := \mathit{CofA}(p_A); \\
\quad \quad \mathit{from} \longleftarrow \mathit{operation}_C(\mathit{to}); \\
\quad \quad r_A := \mathit{AofC}(\mathit{from}) \\
\quad \mathbf{END} \\
\mathbf{END}
\end{array}
$$
\caption{Schematic refinement that accommodates signature changes}
\label{fig:refinement-pattern}
\end{figure}

The conversion functions between $\mathit{type}_A$ and $\mathit{type}_C$
are declared and specified in the component $\mathit{Context}_I$, shown in Figure \ref{fig:context-interface}.
The first two properties define the conversion functions $\mathit{AofC}$ 
and $\mathit{CofA}$ as total bijective functions. The third property 
constrains that they inverse each other. The properties numbered 4 to 7
constrain the translation functions to preserve the invariant states,
the initial states, and the legal operation parameter values. The
properties 8 to 11 further constrain that they preserve the state
transition and output relations.

Atelier~B~\cite{atelierb}, an IDE for the B method, has been used to
develop this pattern.  To show the correctness of the development with
the provers of Atelier-B, we introduced (and proved) the properties
listed in ``assertions'' section.


\begin{figure}
$$
\begin{array}{lr}
\mathbf{MACHINE} \quad \mathit{Context}_I & \\
\mathbf{SEES} \quad \mathit{Context}_A, \mathit{Context}_C & \\
\mathbf{CONSTANTS} \quad \mathit{AofC}, \mathit{CofA} & \\
\mathbf{PROPERTIES} & \\
\quad \mathit{AofC} \in \mathit{type}_C \bij \mathit{type}_A \quad \land & \mathtt{1} \\
\quad \mathit{CofA} \in \mathit{type}_A \bij \mathit{type}_C \quad \land & \mathtt{2} \\
\quad \mathit{CofA}^{-1} = \mathit{AofC}  \quad \land & \mathtt{3}\\
\quad \forall a \bullet (a \subseteq \mathit{type}_A \land a \subseteq inv_A \Rightarrow \mathit{CofA}\lbrack a\rbrack  \subseteq \mathit{inv}_C) \quad \land & \mathtt{4} \\
\quad \forall c \bullet (c \subseteq \mathit{type}_C \land c \subseteq \mathit{inv}_C \Rightarrow \mathit{AofC}\lbrack c\rbrack  \subseteq \mathit{inv}_A) \quad \land & \mathtt{5} \\
\quad \forall c \bullet (c \subseteq \mathit{type}_C \land c \subseteq \mathit{init}_C \Rightarrow \mathit{AofC}\lbrack c\rbrack  \subseteq \mathit{init}_A) \quad \land & \mathtt{6} \\
\quad \forall v_a, p_a \bullet (v_a \subseteq \mathit{type}_A \land p_a \subseteq \mathit{type}_A \land v_a \times p_a \subseteq \mathit{pre}_A \Rightarrow \mathit{CofA}\lbrack v_a \times p_a \rbrack  \subseteq \mathit{pre}_C) \quad \land & \mathtt{7}\\
\quad \forall v, p \bullet (v \in \mathit{type}_A \land p \in \mathit{type}_A \land (v, p) \in \mathbf{dom}(\mathit{stf}_A) \Rightarrow & \mathtt{8}\\
\quad \quad (\mathit{CofA}(v), \mathit{CofA}(p)) \in \mathbf{dom}(\mathit{stf}_C)) \quad \land & \mathtt{}\\
\quad \forall v, p \bullet (v \in \mathit{type}_A \land p \in \mathit{type}_A \land (v, p) \in \mathbf{dom}(\mathit{stf}_A) \Rightarrow & \mathtt{9}\\
\quad \quad \mathit{CofA}\lbrack \mathit{stf}_A\lbrack \{(v, p)\}\rbrack \rbrack  = \mathit{stf}_C\lbrack \{(\mathit{CofA}(v), \mathit{CofA}(p))\}\rbrack ) \quad \land & \\
\quad \forall v, p \bullet (v \in \mathit{type}_A \land p \in \mathit{type}_A \land (v, p) \in \mathbf{dom}(\mathit{ouf}_A) \Rightarrow & \mathtt{10}\\
\quad \quad (\mathit{CofA}(v), \mathit{CofA}(p)) \in \mathbf{dom}(\mathit{ouf}_C))  \quad \land & \\
\quad \forall v, p \bullet (v \in \mathit{type}_A \land p \in \mathit{type}_A \land (v, p) \in \mathbf{dom}(\mathit{ouf}_A) \Rightarrow & \mathtt{11} \\
\quad \quad \mathit{CofA}\lbrack \mathit{ouf}_A\lbrack \{(v, p)\}\rbrack \rbrack  = \mathit{ouf}_C\lbrack \{(\mathit{CofA}(v), \mathit{CofA}(p))\}\rbrack ) & \\
\mathbf{ASSERTIONS} & \\
\quad \mathit{AofC}^{-1} = \mathit{CofA} \quad \land & \\
\quad \mathbf{dom}(\mathit{AofC}) = \mathit{type}_C \quad \land & \\
\quad \mathbf{dom}(\mathit{CofA}) = \mathit{type}_A \quad \land & \\
\quad \forall a, c \bullet (a \in \mathit{type}_A \land c \in \mathit{type}_C \Rightarrow ((\mathit{AofC}(c) = a) \Leftrightarrow (c = \mathit{CofA}(a)))) \quad \land & \\
\quad \forall v_A, p_A \bullet (v_A \in \mathit{type}_A \land p_A \in \mathit{type}_A \land (v_A, p_A) \in \mathit{inv}_A \lhd \mathit{pre}_A \Rightarrow & \\
\quad \quad \mathit{CofA}\lbrack \mathit{stf}_A\lbrack \{(v_A, p_A)\}\rbrack \rbrack  \subseteq \mathit{inv}_C) \quad \land & \\
\quad \forall s \bullet (s \subseteq \mathit{type}_A \Rightarrow \mathit{AofC}\lbrack \mathit{CofA}\lbrack s\rbrack \rbrack  = s) \quad \land & \\
\quad \forall s \bullet (s \subseteq \mathit{type}_C \Rightarrow \mathit{CofA}\lbrack \mathit{AofC}\lbrack s\rbrack \rbrack  = s) & \\
\mathbf{END}& 
\end{array}
$$
\caption{Constraints to establish the refinement pattern for signature changes}
\label{fig:context-interface}
\end{figure}

\section{Case study}
\label{sec:case-study}

In this section, we apply the refinement pattern described in
Section~\ref{sec:interface} in the formal development of a Java Card
implementation of the Counter specification presented in
Section~\ref{sec:counter} and contrast it to the retrenchment approach
exposed in Section~\ref{sec:retsample}.

As seen, a change in the interface of the operations is required, and
in this section the refinement pattern of Section~\ref{sec:interface}
is applied.

The \emph{JCCounter} machine (Figure~\ref{fig:jccounter}) provides the
same services as the \emph{JCounter} machine, but with its interface
and typing restrictions compatible with the types of Java Card. In Java
Card, the type \emph{int} is not built-in and needs to be programmed,
e.g. as a pair of short integers. This representation is defined and
named by \emph{JCINT} in a library machine called \emph{JCInt} (not
detailed in this paper).  Note that the machine \emph{JCCounter} is
also the initial model of a B development to provide an implementation
of the card-side component.

\begin{figure}[ht!]
\begin{small}
\bf MACHINE \hspace*{0.20in}\it JCCounter

\bf SEES \hspace*{0.20in}\it JCInt

\bf VARIABLES \hspace*{0.20in}\it jc\_value
              \hspace*{0.20in}

\bf INVARIANT \hspace*{0.20in}\it jc\_value  $\in$  \it JCINT
              \hspace*{0.20in}

\bf INITIALISATION \hspace*{0.20in}\it jc\_value \rm := \it jcint\_of\rm (\rm 0\rm )


\bf OPERATIONS

\hspace*{0.20in}\bf jc\_increment \rm (\it vv\rm )\rm = 

\hspace*{0.20in}\bf PRE \hspace*{0.20in}\it vv  $\in$  \it JCINT  $\land$ \it sum\_jcint\rm (\it jc\_value\rm , \it vv\rm )  $\in$  \it JCINT

\hspace*{0.20in}\bf THEN \hspace*{0.20in}\it jc\_value  \rm := \it sum\_jcint\rm (\it jc\_value\rm , \it vv\rm )

\hspace*{0.20in}\bf END\rm ;
...






\hspace*{0.20in}\it cc  $\leftarrow$  \bf jc\_getCounterValue \rm =

\hspace*{0.40in}\it cc \rm := \it jc\_value


\bf END
\end{small}
\caption{The JCCounter machine}
\label{fig:jccounter}
\end{figure}

The functions mapping the values of the abstract (Java) and concrete
(Java Card) types are defined in the \emph{InterfaceContext} machine
(see Figure~\ref{fig:interfaceint}). This machine also contains some
corollaries in the assertions clause. These additional properties are
useful to simplify interactive proofs of the development.
These functions are essential to instantiate the refinement pattern to
\emph{JCounter}.

\begin{figure}[ht!]
\begin{small}

\bf MACHINE \hspace*{0.20in}\it InterfaceContext

\bf SEES \hspace*{0.20in}\it JInt\rm ,
\hspace*{0.20in}\it JCInt

\bf CONCRETE\_CONSTANTS \hspace*{0.20in} \it jint\_of\_jcint\rm , \it jcint\_of\_jint \rm

\bf PROPERTIES

\hspace*{0.20in}\it jint\_of\_jcint  $\in$ 

\hspace*{0.40in}\it JCINT  $\pfun$  \it JINT  $\land$ 

\hspace*{0.40in}\it jint\_of\_jcint \rm =  $\lambda$ \rm (\it hi\rm , \it lo\rm )\rm .\rm ( (\it hi \rm, \it lo \rm)   $\in$  \it JCINT $\mid$  \it hi  $\times$  \rm 65536 \rm + \it lo\rm )  $\land$  


\hspace*{0.20in}\it jcint\_of\_jint  $\in$ 

\hspace*{0.40in}\it JINT  $\pfun$  \it JCINT  $\land$ 

\hspace*{0.40in}\it jcint\_of\_jint \rm =  $\lambda$ \rm (\it ii\rm )\rm .\rm (\it ii  $\in$  \it JINT  $\mid$  \rm (\rm (\it ii  $\div$  \rm 65536 )\rm, \rm (\it ii  $\mod$  \rm 6\rm 5\rm 5\rm 3\rm 6\rm )\rm )\rm )\\

\bf ASSERTIONS

\hspace*{0.30in}\it jint\_of\_jcint  $^{-1}$ \rm = \it jcint\_of\_jint  $\land$ 

\hspace*{0.30in}\bf dom \rm (\it jint\_of\_jcint\rm ) \rm = \it JCINT  $\land$ 

\hspace*{0.30in}\bf dom \rm (\it jcint\_of\_jint\rm ) \rm = \it JINT

\hspace*{0.20in}

\bf END

\end{small}
\label{fig:interfaceint}
\caption{The InterfaceContext machine}
\end{figure}

Finally, as a last step, the refinement itself, called
\emph{JCCounter\_ref}, is also obtained by instantiation of the
pattern and is presented in Figure~\ref{fig:counter_ref}. The
development of this case study was also performed and verified with
Atelier~B~\cite{atelierb}.

\begin{figure}[ht!]
\begin{small}
\bf REFINEMENT \hspace*{0.20in}\it JCounter\_ref

\bf REFINES \hspace*{0.15in}\it JCounter

\bf SEES \hspace*{0.20in} \it JInt\rm , \it JCInt\rm , \it InterfaceContext\rm

\bf INCLUDES \hspace*{0.20in}\it JCCounter

\bf INVARIANT \hspace*{0.20in}\it value \rm = \it jint\_of\_jcint\rm (\it jc\_value\rm )

\bf OPERATIONS

\hspace*{0.15in}\bf increment \rm ( \it vv \rm ) \rm =

\hspace*{0.15in}\bf VAR \hspace*{0.20in}\it to

\hspace*{0.15in}\bf IN \hspace*{0.20in}\it to  := \it jcint\_of\_jint\rm (\it vv\rm )\rm;

\hspace*{0.15in}\bf \hspace*{0.15in}\bf jc\_increment\rm (\it to \rm )

\hspace*{0.15in}\bf END;

\hspace*{0.15in} ...

\hspace*{0.15in}\it cc  $\leftarrow$  \bf getCounterValue \rm =

\hspace*{0.15in}\bf VAR \hspace*{0.20in}\it from

\hspace*{0.15in}\bf IN \hspace*{0.20in}\it from  $\leftarrow$  \bf jc\_getCounterValue\rm ;

\hspace*{0.55in}\it cc \rm := \it jint\_of\_jcint\rm (\it from\rm )

\hspace*{0.15in}\bf END

\bf END
\end{small}
\caption{The adapter refinement of counter machine}
\label{fig:counter_ref}
\end{figure}

\section{Conclusions}
\label{sec:conclusions}

The B method provides a simple yet rigorous approach to model-driven
design of software. Starting from an initial functional model of the
requirements, additional requirements and implementation decisions are
introduced as a sequence of refinements. For each refinement, proof
obligations are generated; proving such verification conditions
provides a formal guarantee that the initial specification is indeed an
abstract of model of each successive refinement. 

In the B method, the operations of a refinement must have the same
signature as that of the refined module, and by transitivity, to that
of the initial model. This limitation causes problems in software
developments where component interfaces must be adapted to accommodate,
e.g. incompatibilities in programming languages. 

Retrenchment provides a formal framework to perform model
transformation that is much more flexible than refinement and, in
particular, accommodates interface changes. However one may argue that
the flexibility offered by retrenchment is too generous, to the point
that it may produce implementations that do not conform to the initial
functional specification. Indeed retrenchment is currently not offered
by commercial tools that support the B method. More generally, tool
support for retrenchment as not yet reached the same level of maturity
as refinement.

This paper presents a refinement pattern to accommodate operation
signatures that is fully compatible with the B method. An abstract
instance of this refinement has been developed and verified with
Atelier B~\cite{atelierb}. The paper also shows how the pattern can be
applied in a software development project where different execution
platforms are employed (namely Java and Java Card). This instance has
also been mechanically proved correct.

Future work include:
\begin{enumerate}
\item Proof that the constraints on the interface (or a weaker version
  thereof), listed as properties in the component
  $\mathit{Context}_I$, are necessary conditions to establish the
  refinement. 
\item Automation of the proposed refinement pattern in existing tools
  supporting the B method~\cite{gomes07bsmart} (this includes
  generating verification conditions based on the properties of
  Figure~\ref{fig:context-interface} instead of the more complex
  verification conditions for a generic refinement).
\end{enumerate}
Both lines of work require the construction of an embedding of the B
method in a proof system such as Isabelle~\cite{nipkow05isabelle},
using an approach similar to that of HOL-Z~\cite{brucker03holz}. Such
embedding is necessary to obtain verified results on the B method
(instead of its artifacts as we have done in this paper).

\bibliographystyle{splncs}
\bibliography{bibliography}

\end{document}